\documentclass[runningheads]{llncs}
\usepackage[T1]{fontenc}
\usepackage{graphicx}
\usepackage{amsmath}
\usepackage{amssymb}
\usepackage{booktabs}
\usepackage{wrapfig}
\usepackage{tabularray}
\usepackage[table,xcdraw]{xcolor}
\usepackage{colortbl}
\usepackage[normalem]{ulem}
\useunder{\uline}{\ul}{}
\usepackage{multirow}
\usepackage{pifont}
\usepackage{hyperref}
\usepackage{array}      
\usepackage{siunitx}    
\usepackage{subcaption} 

\usepackage{color}

\makeatletter
\renewcommand{\@fnsymbol}[1]{}
\makeatother

\begin{document}
%


\title{Rethinking the Nested U-Net Approach: Enhancing Biomarker Segmentation with Attention Mechanisms and Multiscale Feature Fusion\thanks{The final version is published in the \textit{Proceedings of the 2024 International Conference on Medical Imaging and Computer-Aided Diagnosis (MICAD 2024)}, Lecture Notes in Electrical Engineering (LNEE), volume 1372, published by Springer Nature. Available at: \href{https://doi.org/10.1007/978-981-96-3863-5_17}{https://doi.org/10.1007/978-981-96-3863-5\_17}}}
\titlerunning{ReN-UNet for Biomarker Segmentation}
%
\author{Saad Wazir \orcidID{0000-0001-9260-1636} \and
Daeyoung Kim \orcidID{0000-0002-7960-5955} }
\authorrunning{}
%
\institute{Korea Advanced Institute of Science and Technology (KAIST), Daejeon, South Korea
\email{\{saad.wazir,kimd\}@kaist.ac.kr}}

\maketitle              
\begin{abstract}
Identifying biomarkers in medical images is vital for a wide range of biotech applications. However, recent Transformer and CNN based methods often struggle with variations in morphology and staining, which limits their feature extraction capabilities. In medical image segmentation, where data samples are often limited, state-of-the-art (SOTA) methods improve accuracy by using pre-trained encoders, while end-to-end approaches typically fall short due to difficulties in transferring multiscale features effectively between encoders and decoders. To handle these challenges, we introduce a nested UNet architecture that captures both local and global context through Multiscale Feature Fusion and Attention Mechanisms. This design improves feature integration from encoders, highlights key channels and regions, and restores spatial details to enhance segmentation performance. Our method surpasses SOTA approaches, as evidenced by experiments across four datasets and detailed ablation studies. Code: \href{https://github.com/saadwazir/ReN-UNet}{https://github.com/saadwazir/ReN-UNet}

\end{abstract}

\section{Introduction} Segmenting biomarkers in medical images is critical for biomedical applications, as it aids in precise diagnoses and guides surgical procedures. This process, including gland and nuclei segmentation, faces challenges such as limited sample availability, which complicates the application of semantic segmentation. CNN-based architectures like UNet \cite{inp:17} and its advanced versions (nested \cite{a:5} and recurrent \cite{a:25}) have enhanced segmentation performance. However, they struggle to capture long-range relationships. Methods like dilated convolutions \cite{a:31} and attention mechanisms \cite{a:6} offer partial solutions, but CNNs' reliance on localized receptive fields limits their ability to capture broader dependencies effectively \cite{a:50}, resulting in only incremental improvements. Conversely, Vision Transformers excel at capturing long-range dependencies but often lack precision in modeling local context. This limitation has been addressed by hybrid models that combine convolutional layers with transformers \cite{inp:39}, effectively managing both local and global contextual information. However, transformer-based methods generally require large datasets, and their self-attention mechanism limits learning fine-grained local pixel relationships \cite{a:48}. Consequently, further improvements are needed to fully capture pixel dependencies across all spatial dimensions

Despite advancements in CNN and transformer-based methods, there is still a need for approaches that fully capture pixel dependencies across all spatial dimensions and accurately segment boundaries. Our end-to-end nested UNet, equipped with novel attention mechanisms, bridges this gap without relying on pre-trained networks. Notably, it surpasses SOTA methods even on smaller datasets, delivering superior segmentation accuracy. In this study, our primary focus is on biomarkers like nuclei and mitochondria segmentation in medical images. Our main contributions are as follows:
\begin{itemize}
    \item We proposed a nested encoder-decoder network designed to efficiently capture long-range, multi-scale contextual information through multiscale feature fusion and channel-wise feature recalibration.
    \item We introduced an Attention Module that enables the network to focus on the most relevant features during upsampling.
    \item An edge enhancement layer and edge-aware loss were developed to place greater emphasis on edge accuracy.
    \item Extensive experiments were conducted on four different segmentation datasets. Our results demonstrate that these advancements collectively improve performance, achieving higher accuracy than the SOTA methods.
\end{itemize}

\section{Related Work}
\textbf{CNN based architectures:} In recent years, extensive research has focused on integrating multi-level deep features for semantic segmentation. Studies \cite{inp:10} have shown that incorporating features from multiple deep layers improves segmentation performance. Variants of UNet, such as U2-Net \cite{a:5} and UNet++ \cite{inp:15}, leverage nested UNet structure and multi-scale feature fusion to capture both local and global context. These models also employ nested and dense skip connections to enhance the integration of feature maps between the encoder and decoder, and incorporate deep supervision to further boost decoder performance. To some extent, attention-based approaches like MA-Unet \cite{inp:21} address long-range dependencies by combining multi-scale features with attention mechanisms. Similarly, Attention U-Net \cite{a:6} applies spatial attention, while Residual-Attention UNet \cite{inp:22} introduces residual attention for more refined feature learning. The nnU-Net \cite{a:34} further enhances adaptability by incorporating self-tuning preprocessing, allowing the model to optimize predictions across different datasets. Despite improvements, these method often need greater adaptability to effectively capture global context and to perform optimally across various data sets \cite{a:50}.

\textbf{Transformer-based architectures:} In image segmentation, Transunet \cite{a:21} incorporates a transformer-driven encoder for robust feature extraction, paired with a U-Net-like decoder to efficiently upsample features. The Swin-unet \cite{inp:9} adopts a hierarchical Swin Transformer encoder with shifted windows for capturing context, combined with a symmetric decoder and a patch-expansion layer to effectively recover spatial resolution. UCTransNet \cite{inp:36} implements channel-wise cross-fusion and attention mechanisms within skip connections, built on a transformer backbone. However, these methods focus on maximizing existing backbone features and heavily rely on large datasets to capture long-range dependencies, while the self-attention mechanism limits their ability to learn fine-grained local pixel relationships \cite{a:48}.

\section{Method}
\begin{figure}[h!]
\centerline{\includegraphics[width=1\textwidth]{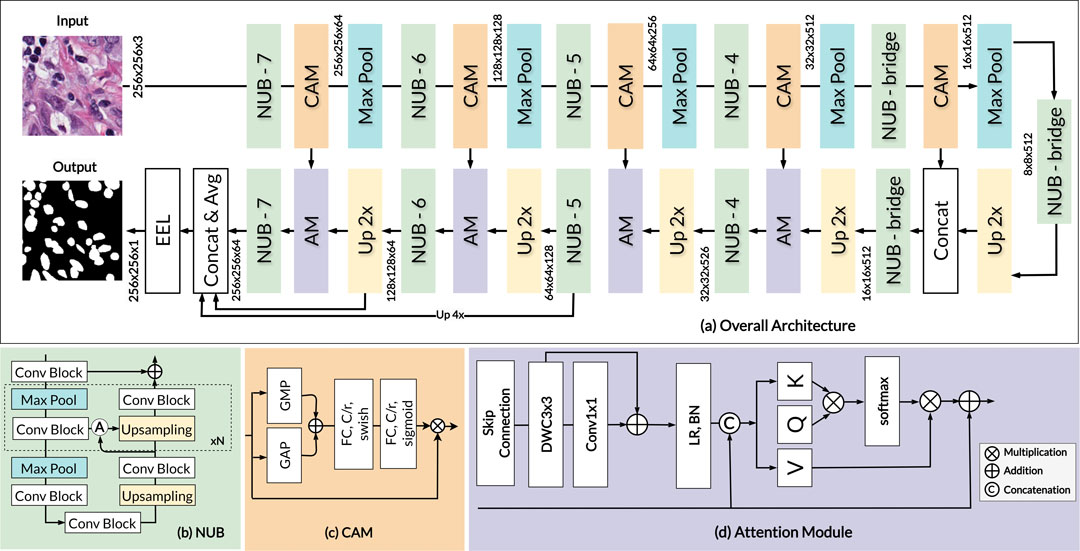}}
\caption{(a) An overview of the proposed architecture (b) Nested UNet Block (NUB). (c) Channel Attention Module (CAM). (d) Attention Module (AM).}
\label{fig:arch}
\end{figure}

\subsection{Overview}
The architecture proposed in this study can be described as a Nested UNet structure, consisting of multiple inner-UNets and a single outer-UNet. In this section, we will discuss the various building blocks and conclude with an explanation of the overall architecture.

\subsection{Building Blocks}
\begin{enumerate}

\item \textbf{Convolution Block:} The convolution block includes depthwise separable convolution layers, batch normalization, and LeakyReLU activation to learn and refine feature representation and introduce non-linearity.

\item \textbf{Nested UNet Block (NUB)} 
We propose a NUB, as illustrated in Figure \ref{fig:arch} (b). The UNet-style encoder-decoder structure, with attention gate \cite{a:6} followed by residual connection, helps capture long-range, multi-scale features with rich contextual information.
We propose variants of the NUB, integrated into the overall architecture, denoted as NUB-N. Here, N represents the sum of the first convolution block and the total number of convolution blocks in the encoder. (N-1) refers to the number of convolution blocks in the encoder, (N-2) indicates the number of max-pooling layers in the encoder and upsampling layers in the decoder, and (N+N) represents the total number of convolution blocks in the NUB. These variants maintain symmetrical configurations with attention gates and skip connections. Specifically, we propose NUB-7, NUB-6, NUB-5, NUB-4, and NUB-bridge. NUB-bridge is similar to NUB-4, except that convolutional blocks are replaced with dilated convolutions.

\item \textbf{Channel Attention Module (CAM):} We introduce CAM, as depicted in Figure \ref{fig:arch} (c), designed to selectively focus on the most significant channels by capturing global channel relationships, enabling the model to effectively learn both global and local patterns. The process starts with feature refinement, followed by global average pooling (GAP) and global max pooling (GMP), which capture the overall feature context and key activations, respectively. These pooled features are combined and passed through a dense layer, utilizing Swish activation \cite{a:43} for non-linearity and smoother gradient propagation. Finally, a dense layer with Sigmoid activation produces attention weights that adjust the original feature map, emphasizing the most important channels.

\item \textbf{Attention Module (AM):} To emphasize essential features from the skip connections, we introduce an attention mechanism that efficiently integrates two feature maps: one from the deeper layers, which captures more comprehensive feature representations, and another from the earlier layers, which maintains finer spatial details. The attention mechanism, as depicted in Figure \ref{fig:arch} (d), refines the skip connections by applying a residual convolution, followed by concatenation with the decoder’s feature map. After this the feature map is transformed into keys, queries, and values using dense layers. The keys and queries calculate attention weights through a dot product, which are then normalized by the softmax function. These weights highlight the most important regions of the feature map, enabling the network to concentrate on crucial information. The weighted values are then combined with the original feature map via a residual connection, preserving the decoder’s features while enriching them with attention.

\item \textbf{Edge Enhancement Layer (EEL):} The Sobel filters \cite{a:35} for horizontal ($S_x$) and vertical ($S_y$) edge detection are applied to the input feature maps to detect changes in intensity in the horizontal and vertical directions, respectively.
Let $I$ be the input feature map to the edge enhancement layer, where $I: \mathbb{R}^{H \times W \times C} \rightarrow \mathbb{R}$, with $H$, $W$, and $C$ denoting the height, width, and channel depth of the feature map, respectively.
The depthwise convolution operations apply the Sobel filters to each channel of $I$, producing edge maps $E_x$ and $E_y$ for horizontal and vertical edges, respectively. The operations can be described as $E_x = I * S_x$ and $E_y = I * S_y$, where $*$ denotes the convolution operation.
The final enhanced edge map $E$ is computed by combining $E_x$ and $E_y$ using the Euclidean norm: $E = \sqrt{E_x^2 + E_y^2}$.
\end{enumerate}

\subsection{Overall Architecture.}
The overall architecture, as shown in Figure \ref{fig:arch} (a), integrates all the components described above. Its primary goal is to efficiently extract, merge, and upsample features. The encoder consists of five NUBs, each followed by a CAM for channel-wise recalibration and max pooling. The Bridge Block refines the encoded features. The decoder mirrors the encoder, progressively restoring spatial dimensions, while utilizing attention mechanisms and skip connections for accurate feature fusion. Each upsampling layer is followed by an NUB to generate multi-scale, feature-rich masks. A pixel-wise mean is calculated in the end, and the result is passed through an edge enhancement layer for refinement. Finally, a sigmoid activation is applied to produce the binary mask. This approach allows the network to capture diverse spatial features and combine them effectively to produce masks that encompass a wide range of scales and contexts. This versatile Nested UNet architecture effectively captures intricate patterns, highlights essential information, and produces accurate predictions for biomarker segmentation in medical images.

\section{Experiments}
\subsection{Datasets and Evaluation Metrics}
To assess our approach, we utilized four publicly available segmentation datasets: Multi-organ Nucleus Segmentation (MoNuSeg) \cite{a:1}, the 2018 Data Science Bowl (DSB) \cite{a:3}, Triple-negative breast cancer (TNBC) \cite{a:2}, and Electron Microscopy (EM) \cite{inp:24}. These datasets include images captured under varying lighting conditions and magnifications, featuring different cell types and imaging modalities. We generated 256x256 overlapping patches with a stride of 128 for all training sets. For testing, we created similar patches, made predictions, reassembled them, and assessed the results. To compare SOTA methods with our proposed method, we employed standard metrics \cite{a:12}. For pixel-based evaluations, we used Intersection over Union (IoU), Dice coefficient, precision, recall, and false omission rate. For surface distance metrics, we calculated the 95th percentile of the Hausdorff distance (HD95) and the average surface distance (ASD), providing a thorough assessment. All metrics were averaged across all samples.

\begin{table}[htbp]
\centering

\centering
\caption{Quantitative comparison on the MoNuSeg Dataset. The best performance is highlighted in bold, and the second-best is underlined.}
\vspace{0.2cm}
\label{tab:monuseg}
\resizebox{\textwidth}{!}{%
\begin{tabular}{llllllll}
\hline
Model      & \multicolumn{1}{c}{IoU $\uparrow$} & \multicolumn{1}{c}{Dice $\uparrow$} & \multicolumn{1}{c}{Prec. $\uparrow$} & \multicolumn{1}{c}{Rec. $\uparrow$} & \multicolumn{1}{c}{FOR $\downarrow$} & \multicolumn{1}{c}{HD95 $\downarrow$} & \multicolumn{1}{c}{ASD $\downarrow$} \\ \hline
U-Net      & 59.62                   & 73.42                    & 74.87                     & 73.04                    & 0.0766                  & 3.6609                   & 0.8109                  \\
Hover-Net  & 62.90                   & 77.26                    & 77.67                     & 77.54                    & 0.0689                  & 3.8687                   & 0.8114                  \\
nnUNet     & 67.60                   & 80.42                    & \textbf{81.63}            & 80.88                    & 0.0555                  & 3.3357                   & 0.6852                  \\
UNet++     & \underline{69.34}       & \underline{81.83}        & 75.31                     & \textbf{90.29}           & 0.0422                  & \underline{2.7901}       & \underline{0.6458}      \\ 
U2-Net     & 68.89                   & 81.33                    & 77.20                     & 86.88                    & 0.0444                  & 13.445                   & 3.0647                  \\
TransUNet  & 67.89                   & 80.22                    & \underline{81.58}         & 79.17                    & 0.0610                  & 3.5602                   & 0.7379                  \\
Swin-Unet  & 65.38                   & 79.01                    & 72.38                     & \underline{87.88}        & \underline{0.0403}      & 3.7871                   & 0.7893                  \\
UCTransNet & 65.56                   & 79.20                    & 75.83                     & 82.88                    & 0.0561                  & 18.366                   & 4.2000                  \\ \hline
Ours       & \textbf{73.06}          & \textbf{84.12}           & 81.46                     & 87.30                    & \textbf{0.0329}         & \textbf{2.2422}          & \textbf{0.1583}         \\ \hline
\end{tabular}%
}
\end{table}

\begin{table}[htbp]
\centering
\caption{Quantitative comparison on the DSB 2018 Dataset.}
\vspace{0.2cm}
\label{tab:dsb}
\resizebox{\textwidth}{!}{%
\begin{tabular}{llllllll}
\hline
Model      & \multicolumn{1}{c}{IoU $\uparrow$} & \multicolumn{1}{c}{Dice $\uparrow$} & \multicolumn{1}{c}{Prec. $\uparrow$} & \multicolumn{1}{c}{Rec. $\uparrow$} & \multicolumn{1}{c}{FOR $\downarrow$} & \multicolumn{1}{c}{HD95 $\downarrow$} & \multicolumn{1}{c}{ASD $\downarrow$} \\ \hline
U-Net      & 83.83                   & 90.17                    & 89.09                     & 93.78                    & 0.0128                  & 7.6797                   & \underline{1.9192}      \\
Hover-Net  & 81.29                   & 89.04                    & 88.73                     & 91.01                    & 0.0246                  & \underline{7.6289}       & 2.1298                  \\
nnUNet     & 77.24                   & 86.09                    & 90.18                     & 84.33                    & 0.0337                  & 9.9094                   & 2.7911                  \\
UNet++     & 83.95                   & 91.10                    & 88.49                     & \underline{94.26}        & \underline{0.0115}      & 10.045                   & 2.4359                  \\
U2-Net     & 82.96                   & 90.03                    & 90.28                     & 91.08                    & 0.0163                  & 10.869                   & 2.3375                  \\
TransUNet  & 84.72                   & 90.90                    & \underline{93.37}         & 89.86                    & 0.0162                  & 11.213                   & 2.7168                  \\
Swin-Unet  & 84.49                   & 91.03                    & 90.77                     & 92.72                    & 0.0132                  & 8.0415                   & 1.9232                  \\
UCTransNet & \underline{85.28}       & \underline{91.25}        & \textbf{93.87}            & 90.02                    & 0.0158                  & 11.514                   & 2.5552                  \\ \hline
Ours       & \textbf{87.22}          & \textbf{92.79}           & 91.79                     & \textbf{94.76}           & \textbf{0.0032}         & \textbf{6.5914}          & \textbf{1.7074}         \\ \hline
\end{tabular}%
}
\end{table}

\begin{table}[htbp]
\centering

\centering
\caption{Quantitative comparison on the Electron Microscopy Dataset.}
\vspace{0.2cm}
\label{tab:electron}
\resizebox{\textwidth}{!}{%
\begin{tabular}{llllllll}
\hline
Model      & \multicolumn{1}{c}{IoU $\uparrow$} & \multicolumn{1}{c}{Dice $\uparrow$} & \multicolumn{1}{c}{Prec. $\uparrow$} & \multicolumn{1}{c}{Rec. $\uparrow$} & \multicolumn{1}{c}{FOR $\downarrow$} & \multicolumn{1}{c}{HD95 $\downarrow$} & \multicolumn{1}{c}{ASD $\downarrow$} \\ \hline
U-Net      & 77.45                   & 85.37                    & 88.82                     & 84.95                    & 0.0100                  & 46.803                   & 10.658                  \\
Hover-Net  & 79.18                   & 88.31                    & 91.43                     & 85.55                    & 0.0084                  & 12.624                   & 2.1152                  \\
nnUNet     & 85.02                   & 91.86                    & \textbf{94.75}            & 89.23                    & 0.0064                  & \underline{5.2735}       & 1.2231                  \\
UNet++     & \underline{87.01}       & \underline{92.49}        & 92.53                     & \underline{93.38}        & 0.0042                  & \textbf{4.7633}          & \underline{1.1553}      \\
U2-Net     & 86.31                   & 92.01                    & 91.92                     & 92.88                    & \underline{0.0041}      & 20.817                   & 4.3750                  \\
TransUNet  & 84.86                   & 91.77                    & \underline{94.33}         & 89.42                    & 0.0062                  & 6.1746                   & 1.4101                  \\
Swin-Unet  & 83.72                   & 91.08                    & 92.18                     & 90.10                    & 0.0059                  & 9.7707                   & 1.8287                  \\
UCTransNet & 85.99                   & 92.32                    & 91.55                     & 93.24                    & 0.0046                  & 6.7678                   & 1.5115                  \\ \hline
Ours       & \textbf{87.93}          & \textbf{93.55}           & 92.86                     & \textbf{94.37}           & \textbf{0.0028}         & 5.3703                   & \textbf{0.3047}         \\ \hline
\end{tabular}%
}
\end{table}

\begin{table}[htbp]
\centering
\caption{Quantitative comparison on the TNBC Dataset.}
\vspace{0.2cm}
\label{tab:tnbc}
\resizebox{\textwidth}{!}{%
\begin{tabular}{llllllll}
\hline
Model      & \multicolumn{1}{c}{IoU $\uparrow$} & \multicolumn{1}{c}{Dice} & \multicolumn{1}{c}{Prec.} & \multicolumn{1}{c}{Rec.} & \multicolumn{1}{c}{FOR $\downarrow$} & \multicolumn{1}{c}{HD95 $\downarrow$} & \multicolumn{1}{c}{ASD $\downarrow$} \\ \hline
U-Net      & 53.41                   & 67.55                    & \textbf{88.79}            & 58.44                    & 0.0548                  & 21.101                   & 5.0472                  \\
Hover-Net  & 54.89                   & 69.20                    & \underline{88.68}         & 59.57                    & 0.0539                  & 23.017                   & 5.0716                  \\
nnUNet     & 59.35                   & 73.06                    & 84.57                     & 67.63                    & 0.0511                  & 16.768                   & 3.3843                  \\
UNet++     & 60.94                   & 74.53                    & 77.89                     & \underline{75.20}        & 0.0370                  & 17.106                   & 3.9056                  \\
U2-Net     & \underline{61.86}       & \underline{75.63}        & 72.02                     & \textbf{82.43}           & \underline{0.0349}      & \underline{14.659}       & 3.4034                  \\
TransUNet  & 59.71                   & 73.31                    & 85.69                     & 66.90                    & 0.0518                  & 15.273                   & \underline{3.2620}      \\
Swin-Unet  & 55.87                   & 70.20                    & 86.78                     & 62.13                    & 0.0515                  & 19.421                   & 4.3773                  \\
UCTransNet & 56.62                   & 71.29                    & 86.83                     & 62.90                    & 0.0549                  & 18.340                   & 3.5854                  \\ \hline
Ours       & \textbf{66.13}          & \textbf{78.99}           & 86.09                     & 74.97                    & \textbf{0.0291}         & \textbf{10.355}          & \textbf{3.2480}         \\ \hline
\end{tabular}%
}
\end{table}

\subsection{Edge-Aware loss (EAL) }
The EAL function is tailored for image segmentation tasks, aiming to enhance the model's accuracy around the edges of segmented objects. The function begins by calculating the binary cross-entropy (BCE) between the predicted masks \( y_{\text{pred}} \) and the ground truth masks \( y_{\text{true}} \). It then employs Sobel filters, denoted as \( E_x \) and \( E_y \), to detect edges in the ground truth. These filters compute the gradient in the x and y directions respectively, and the magnitude of these gradients is used to create an edge mask. This mask identifies significant edges by thresholding the gradient magnitudes above 0.1. Regions identified as edges are given additional emphasis by multiplying their loss values by \( w \) (where \( w \) is a specified weight greater than 1), enhancing the model's focus on these critical areas. The final loss \( L \) is the mean of all the weighted BCE values across the dataset, ensuring that the loss is averaged over all samples. Edge-aware loss function is defined as:
\begin{equation}
L(y_{\text{true}}, y_{\text{pred}}) = \frac{1}{N} \sum_{i=1}^N \left(1 + \left(\text{em}_i \times (w - 1)\right)\right) \times \text{BCE}(y_{\text{true}_i}, y_{\text{pred}_i})
\end{equation}
where
\[
\text{edge\_mask (em)} = \frac{1}{D} \sum_{d=1}^D \mathbf{1}\left(\sqrt{s_x^2 + s_y^2} > 0.1\right),
\]
\[
s_x = \text{Sobel}_x(y_{\text{true}}), \quad s_y = \text{Sobel}_y(y_{\text{true}}),
\]
and \( w \) is the edge weight, \( N \) is the number of samples, and \( D \) is the spatial dimensionality of the mask.

\subsection{Experimental Setup and Results}
We trained the model using the Adam optimizer with a learning rate of 1e-4 and a batch size of 8, applying the HeUniform kernel initializer for the convolution layers. The model was trained from scratch for 200 epochs on the training sets of the MoNuSeg, DSB, and EM datasets, with the best-performing weights selected for evaluation. Both offline and online data augmentation techniques were employed. To ensure a fair comparison, we reproduced the results for all datasets, averaging them over five runs. The provided test sets for MoNuSeg and EM were used for evaluation, while for DSB, we used a 9:10 split for training and testing. TNBC dataset is only used for evaluation using MoNuSeg weights.

\begin{figure}{}{}
\centerline{\includegraphics[width=0.7\textwidth]{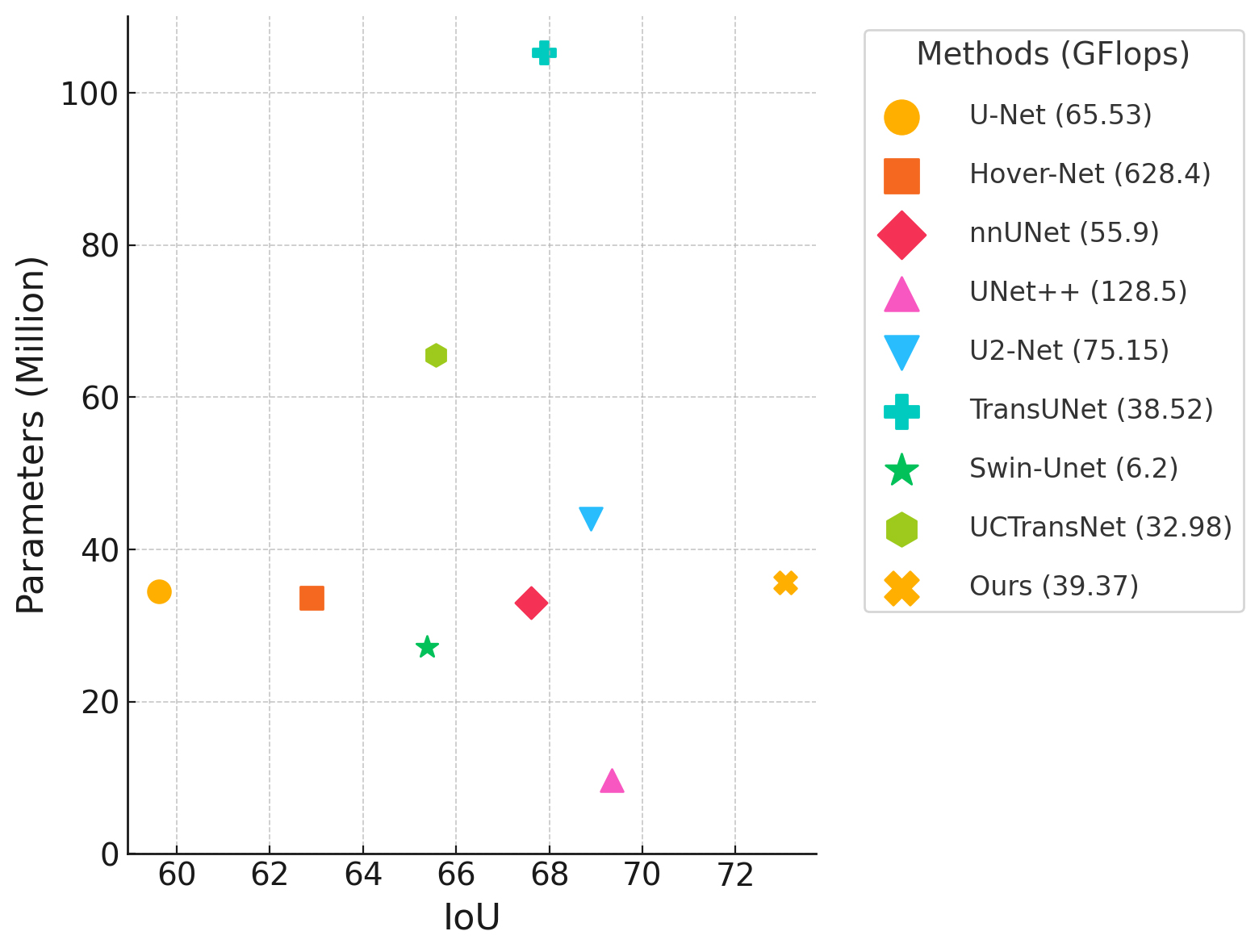}}
\caption{Comparison of Trainable parameters, GFlops, and mIoU across models.}
\label{fig:params}
\end{figure}

\begin{figure}[h!]
\centerline{\includegraphics[width=1\textwidth]{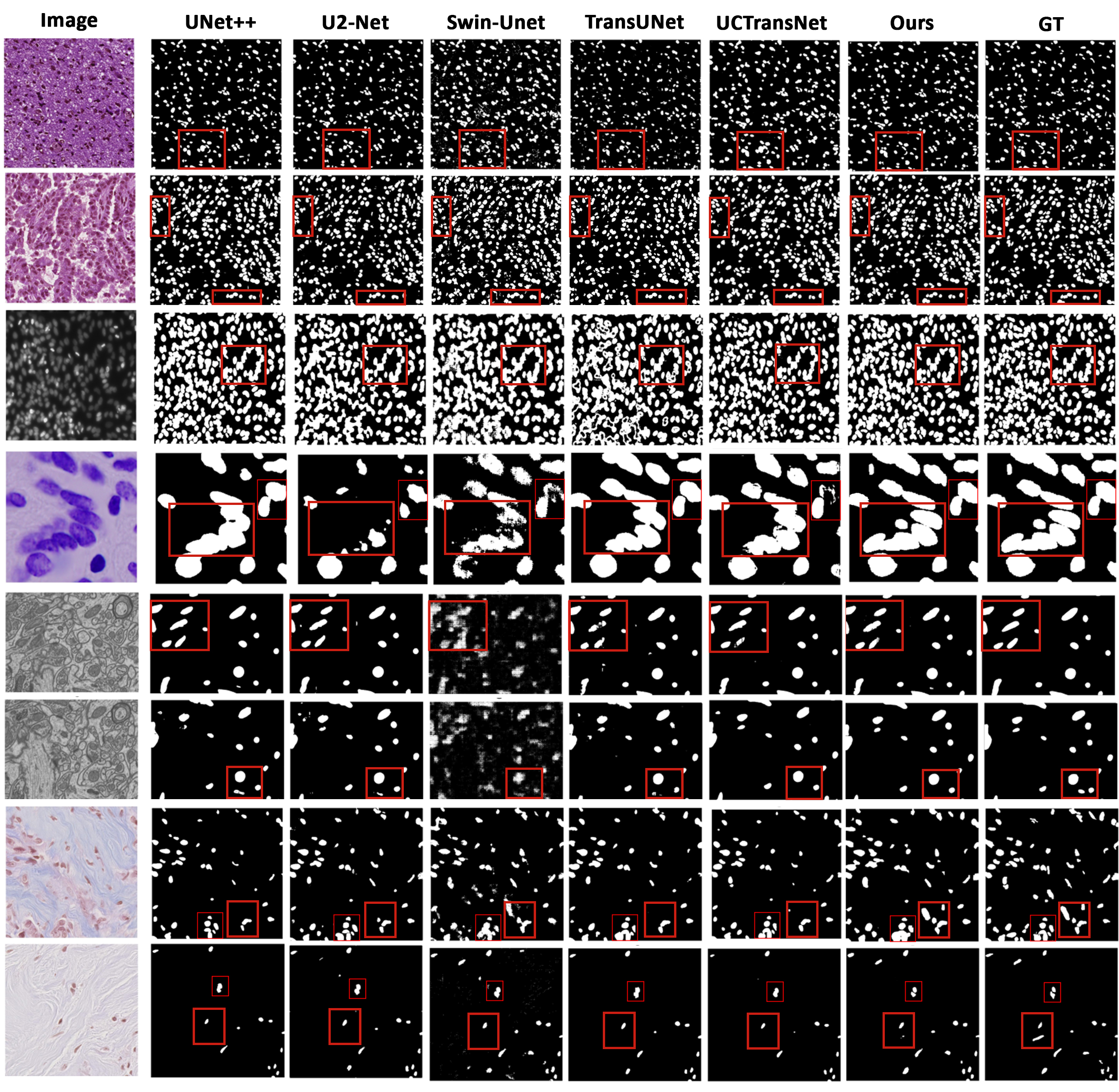}}
\caption{Qualitative Results Comparison: Black pixels represent the background, while white pixels represent the biomarker. The red box indicates the region where there is either no prediction or over-segmentation.}
\label{fig:qual}
\end{figure}

Our evaluation demonstrates that our proposed model leads in major metrics, indicating its robustness for segmenting biomarkers, as shown in Table \ref{tab:monuseg}, \ref{tab:dsb}, \ref{tab:electron}, and \ref{tab:tnbc} . In terms of efficiency, our model has relatively fewer parameters compared to its closest counterparts and nearly one-third the GFLOPs, as shown in Figure \ref{fig:params}. Although our model is less efficient than UNet++ and Swin-Unet, it offers higher accuracy as a trade-off. For the analysis of qualitative results, we selected diverse samples showcasing different modalities. By analyzing the results from each dataset, as shown in Figure \ref{fig:qual}, we can conclude that our method not only performs accurate predictions but also avoids creating background noise and over-segmentation.

\subsection{Ablation Studies}
\label{sec:abl-study}
\begin{table}[htbp]
\centering
\centering
\caption{Design choices and their impacts in the proposed approach.}
\vspace{0.2cm}
\label{tab:abl-design-choice}
\resizebox{0.7\textwidth}{!}{%
\begin{tabular}{@{}lllllllllll@{}}
\toprule
\textbf{in\_A} & \textbf{AM} & \textbf{CAM} & \textbf{EEL} & \textbf{Params} & & \textbf{GFlops} & & \textbf{IoU} & & \textbf{HD95} \\ \midrule
\ding{55}      & \ding{55}   & \ding{55}    & \ding{55}    & 5.5187          &  & 10.32           &  & 66.05        &  & 3.1266        \\
\checkmark     & \ding{55}   & \ding{55}    & \ding{55}    & 12.122          &  & 17.91           &  & 68.28        &  & 3.0010        \\
\ding{55}      & \checkmark  & \ding{55}    & \ding{55}    & 33.001          &  & 27.59           &  & 68.65        &  & 2.7607        \\
\checkmark     & \checkmark  & \ding{55}    & \ding{55}    & 35.650          &  & 38.95           &  & 71.17        &  & 2.5407        \\
\checkmark     & \checkmark  & \checkmark   & \ding{55}    & 35.650          &  & 39.02           &  & 72.68        &  & 2.4161        \\
\checkmark     & \checkmark  & \checkmark   & \checkmark   & 35.650          &  & 39.37           &  & 73.06        &  & 2.2422        \\ \bottomrule
\end{tabular}%
}
\end{table}

\begin{table}[htbp]
\centering
\caption{Comparison of Loss Functions.}
\vspace{0.2cm}
\label{tab:abb-loss}
\resizebox{0.7\textwidth}{!}{%
\begin{tabular}{@{}lccccc@{}}
\toprule
\textbf{Loss} & \textbf{IoU $\uparrow$} & \textbf{Prec. $\uparrow$} & \textbf{Rec. $\uparrow$} & \textbf{Dice $\uparrow$} & \textbf{HD95 $\downarrow$} \\ \midrule
BCE           & 68.70                   & 79.37                      & 83.61                    & 81.39                    & \underline{2.6625}         \\
Dice Loss     & 69.33                   & 77.21                      & \textbf{88.07}           & 81.73                    & 3.2401                     \\
BCE + Dice    & \underline{71.43}       & \textbf{82.84}             & 83.82                    & \underline{83.21}        & 2.6938                     \\
Focal Loss    & 67.88                   & 76.03                      & 86.69                    & 80.81                    & 2.9113                     \\
Ours          & \textbf{73.06}          & \underline{81.46}          & \underline{87.30}        & \textbf{84.12}           & \textbf{2.2422}            \\ \bottomrule
\end{tabular}%
}
\end{table}

We conducted all ablation studies using MoNuSeg dataset to assess the effectiveness of components in our model, as detailed in Table \ref{tab:abl-design-choice}. Our baseline involves nested U-Nets in an encoder-decoder setup and our loss function. we progressively introduced each component, observing improvements across key metrics. Adding inner attention (in\_A) improves the model's ability to focus on important spatial regions, leading to performance gains. AM enhances attention to relevant areas for multi-scale feature integration, CAM sharpens feature representation, and EEL enhances boundary detection. Collectively, these components enhance the model's performance.
When evaluating our proposed loss function, as shown in Table \ref{tab:abb-loss}, we found that BCE performs moderately, while Dice Loss excels in recall by effectively balancing error margins. The combined BCE + Dice Loss offers the best precision, whereas Focal Loss shows weaker performance in IoU, Precision, and HD95, indicating higher boundary errors. Our proposed loss function outperforms the others in most metrics by emphasizing edge regions, leading to more accurate positive identification and better coverage of positive instances. We evaluate the EAL and EEL separately, as shown in Table \ref{tab:abl-eal-eel}. While some models showed slight improvements, others, particularly transformer-based methods, experienced a decrease in performance, which may be due to interference with the models' intrinsic learning mechanisms, as additional edge emphasis can disrupt their already complex architectures and training dynamics.

\begin{table}[htbp]
\centering
\caption{Quantitative Comparison of EAL Function and EEL.}
\vspace{0.2cm}
\label{tab:abl-eal-eel}
\begin{tabular}{lccccccccccccccccccccccccccc}
\hline
\textbf{Model} & & & & & & & & & & & & & & & &   \textbf{IoU} & & & \textbf{IoU (EAL)} & & & \textbf{IoU (EEL)} & & \\ \hline
UNet           & & & & & & & & & & & & & & & & 59.62        & & & 61.35              & & & 63.97              & & \\
UNet++         & & & & & & & & & & & & & & & & 69.34        & & & 69.71              & & & 70.82              & & \\
U2-Net         & & & & & & & & & & & & & & & & 68.89        & & & 69.05              & & & 70.51              & & \\
Transunet      & & & & & & & & & & & & & & & & 67.89        & & & 64.35              & & & 65.09              & & \\
Swin-Unet      & & & & & & & & & & & & & & & & 65.38        & & & 66.42              & & & 66.50              & & \\
UCTransNet     & & & & & & & & & & & & & & & & 65.56        & & & 65.47              & & & 64.92              & & \\ \hline
Ours           & & & & & & & & & & & & & & & & 73.06        & & & -                  & & & -                  & & \\ \hline
\end{tabular}
\end{table}

\section{Conclusion}
Our research introduces an innovative model that effectively captures both local and global contextual information using a nested UNet approach and channel recalibration. We integrate multi-scale features through an Attention Module to extract relevant features and enhance upsampling. Extensive evaluations on four datasets validate our model’s performance. Moving forward, we plan to expand our research to include additional medical image segmentation tasks, such as gland, polyp, and organ segmentation.

\section{Acknowledgments}
This research was supported by the MSIT (Ministry of Science and ICT), Korea, under the ITRC program (RS-2024-00259703) supervised by IITP.

%
%
%
\bibliographystyle{splncs04}
\bibliography{main}

\end{document}